\documentclass[11pt]{article}
\usepackage{geometry}                
\usepackage[parfill]{parskip}
\usepackage{amssymb}

\usepackage{amsmath}

\title{CLASSICAL AND QUANTUM CORRELATION FUNCTIONS FOR A RING MODEL}

\author{B.~Buck and C.~V.~Sukumar \\{\em The Rudolf Peierls Centre for Theoretical Physics,}\\{\em Department of Physics, University of Oxford, Oxford OX1 3NP, U.K. }}

\begin{document}
\maketitle
\begin{abstract} 
Classical and quantum correlation functions are derived for a system of non-interacting particles moving on a circle. It is shown that the decaying behaviour of the classical expression for the correlation function can be recovered from the strictly periodic quantum mechanical expression by taking the limit $\hbar\rightarrow 0$ after an appropriate transformation.
\end{abstract}
\vfill\centerline{PACS INDICES:~~03.65, 05.20, 05.30 and 05.70}
\vfill\eject

\section{INTRODUCTION}
\medskip\noindent 
A model of a non-interacting gas confined to move on the perimeter of circle was considered by Frisch~[1]. In this model it is possible to study the time evolution of many-particle averages such as the local densities of particles and momenta and thereby establish how the various averages approach equilibrium values, starting from specific initial conditions. Later, Hobson and Loomis~[2] considered a variant of the model in which a Knudsen gas was confined to move in a three dimensional rectangular box with the possibility of elastic reflections at the walls.  In both of these calculations, which were performed within the framework of classical Statistical Mechanics, it was found that the many-particle averages of local observables converge to equilibrium values in a qualitatively similar fashion.

In this paper we consider the Frisch model and show that the quantum mechanical equations for the time evolution of various operators can be solved exactly. The correlation functions for these operators can also be constructed. The Poisson summation formula can then be used to show how the seeming conflict between the strictly periodic nature of the quantal solutions and the irreversible decay that characterises the classical solutions can be resolved in a satisfactory way.

\section{RING MODEL}
\medskip\noindent 
The Hamiltonian of a particle of mass $m$ confined in a circular orbit of radius $R$ is
$$H\ =\ \frac{L^2}{2mR^2},$$
where 
$$L\ =\ -i\hbar\ \frac{\partial}{\partial\phi}$$
and $\phi$ is the angular position of the particle.
Starting from the fundamental commutator $$[\phi, L] = i\hbar,$$
it is easy to establish that the operator $E$ defined by $E = e^{i\phi}$ satisfies $[E,L] = -\hbar E$.
The Heisenberg equation of motion for the operator $E$ is then given by 
$$i\hbar{\dot E}\ =\ [E,H]\ =\ -\ \frac{\hbar}{2mR^2}\ (LE\ +\ EL),$$
with the formal solution
$$E(t)\ =\ {e^\frac{itL}{2mR^2}}\ E(0)\ {e^\frac{itL}{2mR^2}}.$$
The operator $E(t)$ can also be written in the alternative forms
$$E(t)\ =\ E(0)\ e^{\frac{it}{2mR^2}(2L + \hbar)}$$
and 
$$E(t)\ =\ e^{\frac{it}{2mR^2}(2L - \hbar)}\ E(0)\ .$$

From these solutions for $E$ and the corresponding solutions for the adjoint operator $E^{\dag}$, expressions for the cartesian position operators $x = R\cos\phi$ and $y =R\sin\phi$ can be derived in the equivalent forms
\begin{align}
x(t)\ &=\ \left[x(0)\ \cos\left(\frac{Lt}{mR^2}\right)\ -\ y(0)\ \sin\left(\frac{Lt}{ mR^2}\right)\right]\ e^{+\frac{i\hbar t}{2mR^2}} \notag\\
 &=\ \left[\cos\left(\frac{Lt}{mR^2}\right)\ x(0)\ -\ \sin\left(\frac{Lt}{mR^2}\right) \ y(0)\right]\ e^{-\frac{i\hbar t}{2mR^2}}, \notag\\ 
y(t)\ &=\ \left[y(0)\ \cos\left(\frac{Lt}{mR^2}\right)\ +\ x(0) \sin\left(\frac{Lt}{ mR^2}\right)\right]\ e^{+\frac{i\hbar t}{2mR^2}} \notag\\
 &=\ \left[\cos\left(\frac{Lt}{mR^2}\right)\ y(0)\ +\ \sin\left(\frac{Lt}{mR^2}\right) \ x(0)\right]\ e^{-\frac{i\hbar t}{2mR^2}}. \label{}
\end{align}
It is clear from these expressions that $x(0)$ does not in general commute with $x(t)$ and similarly for the $y$ operators.

In the ring model we consider $N$ non-interacting point particles of identical mass $m$ moving on a circle of radius $R$. Let $\phi_i$ be the angular position on thecircle of the $i$'th particle. The Hamiltonian of this system is 
$$H\ =\ \frac{1}{2mR^2}\ \sum_i L_i^2$$ 
in which $L_i$ is the angular momentum of the $i$'th particle as  represented by $L_i = -i\hbar\ \frac{\partial}{\partial\phi_i}.$ Since the particles are non-interacting, the Hamiltonian is already separable in the coordinates of the particles. Therefore, in determining properties of the $N$ particle system such as the average energy or the position correlation function for the centre of mass, it is possible to remove the sums over particle labels. By considering just one particle, but allowing position, energy, etc., to be determined by a distribution function or single particle density matrix, the single particle averages can be set equal to the $N$ particle averages. If, at time $t = 0$, the available information on the $N$-particle system consists only of a knowledge of the total mean energy ${\bar E}$, then according to the principle of maximum entropy~[3] the density matrix defined by
\begin{align}
\rho\ &=\ \frac{1}{Z} \ e^{-\beta H},\notag\\
Z\ &=\ {\rm Tr}(e^{-\beta H}),\label{}
\end{align}
with the Lagrange multiplier $\beta$ determined from 
$${\bar E}\ =\ \langle H\rangle _{t = 0}\ =\ {\rm Tr}(\rho H)\ =\ -\ \frac{\partial}{\partial\beta}\ \ln Z,$$ 
maximises the entropy $S = {\rm Tr}(\rho\ln\rho).$
The initial density matrix in this instance is determined completely by the average energy of the particles. A density matrix so constructed encodes all the available initial information in a consistent way. The best estimate for the expectation value of any operator $A(t)$ is then
$${\langle A\rangle}_t\ =\ {\rm Tr}[\rho A(t)]\ .$$

In classical mechanics $A(t)$ is found by solving the appropriate Lagrangian or Hamiltonian equations of motion. In quantum mechanics $A(t)$ is a solution of the Heisenberg equation of motion
$$i\hbar\dot A\ =\ [A,H],$$
and may be written in the form
\begin{equation}
A(t)\ =\ e^{\frac{iHt}{\hbar}}\ A(0)\ e^{-\frac{iHt}{\hbar}}. \label{}
\end{equation}

\section{CORRELATION FUNCTIONS}
\medskip\noindent 
The correlation functions of an operator $A(z)$ with $A(0)$ may be defined by
\begin{align}
C_1(z)\ &=\ \langle A(0)A(z)\rangle\ =\ {\rm Tr}[A(0)A(z)\rho(0)],\notag\\
C_2(z)\ &=\ \langle A(z)A(0)\rangle\ =\ {\rm Tr}[A(z)A(0)\rho(0)],\label{}
\end{align}
where $z$ is a complex time variable.
Using Eqs.(2) and (3), the correlation function $C_1$ may be written in the form
$$C_1(z)\ =\ \frac{{\rm Tr}[\ A(0)\ e^{\frac{izH}{\hbar}}\ A(0)\ e^{-\frac{izH}{\hbar}}\  e^{-{\beta H}}\ ]}{{\rm Tr}[\ e^{-{\beta H}}\ ]}.$$
The property of invariance of the trace operation under cyclic permutation of its arguments may be used to show that the numerator of $C_1(-z)$ takes the forms
$${\rm Tr}[\ A(0)\ e^{-\frac{izH}{\hbar}}\ A(0)\ e^{\frac{izH}{\hbar}}\ e^{-{\beta H}}\ ]\ = \ {\rm Tr}[\ A(0)\ e^{\frac{i(z+i\beta\hbar)H}{\hbar}}\ A(0)\ e^{-\frac{i(z+i\beta\hbar)H}{\hbar}}\ e^{-\beta H}\ ]$$
$$\ =\ {\rm Tr}[\ e^{\frac{izH}{\hbar}}\ A(0)\ e^{-\frac{izH}{\hbar}}\ A(0)\ e^{-\beta H}\ ].$$
Hence we see from Eqs.(4) that 
$$C_1(-z)\ =\ C_1(z + i\beta\hbar)\ =\ C_2(z).$$
The above relations between correlation functions are model independent and are valid whenever the initial density matrix has the canonical form given by Eqs.(2).

The operator solutions for $x(t)$ given in Eqs.(1) may now be inserted in the first of Eqs.(4) to yield the correlation function of $x(0)$ with $x(t)$ when the initial energy is specified, i.e.,
$$C_1(t)\ =\ \langle x(0)\ x(t)\rangle\ =\ \frac{{\rm Tr}[\ x(0)\ x(t)\ e^{-\beta H}\ ]}{{\rm Tr}[\ e^{-\beta H}\ ]}.$$
A convenient basis for evaluation of the trace is the set of orthonormal states
$$\vert n \rangle\ =\ \frac{1}{\sqrt{2\pi}}\ e^{in\phi},~~~~{n = 0, \pm 1,\pm 2,\dots ,}$$
which are eigenstates of $L$ with eigenvalues $\hbar n$ and also eigenstates of $H$ with eigenvalues $E_n ={\hbar^2 n^2 / 2mR^2}$.
The matrix elements given by
$$\langle n\vert\cos^2\phi\vert n\rangle\ =\ \frac{1}{2},$$
$$\langle n\vert\sin\phi \cos\phi\vert n\rangle\ =\ 0,$$
$$\langle n\vert\cos\left(\frac{Lt}{mR^2}\right)\vert n\rangle\ =\ \cos\left(\frac{n\hbar t}{mR^2}\right),$$
$$\langle n\vert\sin\left(\frac{Lt}{ mR^2}\right)\vert n\rangle\ =\ \sin\left(\frac{n\hbar t}{mR^2}\right)$$
and the operator solutions for $x(t)$ given in Eqs.(1) may be used to write $C_1$ in the form
$$C_1(t)\ =\ \frac{R^2}{2}\ \frac{F(t)}{F(0)},$$
where 
$$F(t)\ =\ e^{\frac{it}{2\tau_b}}\ \sum_{n= -\infty}^{\infty}\ e^{-\frac{\tau_a}{ 2\tau_b}n^2}\ \cos\left(\frac{nt}{\tau_b}\right).$$
The time constants $\tau_a$ and $\tau_b$ are defined by
$$\tau_a\ =\ \hbar\beta~~~~~{\rm and}~~~~~\tau_b\ =\ \frac{mR^2}{\hbar}.$$
It is not difficult to show that $F(t)$ may also be written in the form
$$F(t)\ =\ e^{-\frac{\tau_a}{8\tau_b}}\ \sum_{n= -\infty}^{\infty}\ e^{-\frac{\tau_a} {2\tau_b}(n+1/2)^2}\ \cos\left[(n+1/2)\left(\frac{t-i\tau_a /2}{\tau_b}\right)\right].$$
The correlation function $C_2$ is given by
$$C_2(t)\ =\ \langle x(t)x(0)\rangle\ =\ e^{-\frac{it}{\tau_b}}\ C_1(t),$$
and using the above expressions it is easy to verify that
$$C_2(t)\ =\ C_1(-t)\ =\ C_1(t + i\tau_a).$$

The correlation functions $C_1$ and $C_2$ are strictly periodic with period $\tau = 4\pi\tau_b$ and they are also related to a function with an imaginary period. To demonstrate this, consider
$$g(z)\ =\ e^{\frac{z^2}{2\alpha}}\ \sum_{n=-\infty}^{\infty}\ e^{-{\frac{\alpha}{2}\ n^2}}\ \cos nz,$$
which is similar to a Jacobi theta function. When it is written in the form
$$g(z)\ =\ \frac{1}{2}\sum_{n=-\infty}^{\infty}\ \left[\ e^{-{\frac{\alpha}{2}(n-\frac{iz}{\alpha})^2}} \ +\ e^{-{\frac{\alpha}{2}(n+\frac{iz}{\alpha})^2}}\ \right],$$
it is clear that
$$g(z + im\alpha)\ =\ g(z),$$
since the summation index can be shifted by whole integers without altering the value of the function. Thus $g(z)$ is periodic with imaginary period $i\alpha$ and $F(t)$ is related to $g(z)$ by 
$$F(t)\ =\ e^{\frac{iz}{2}}\  e^{-\frac{z^2}{2\alpha}}\ g(z),$$
where $z = \frac{t}{\tau_b}$ and $\alpha = \frac{\tau_a}{\tau_b}$.

We next construct an alternative representation for $F(z)$. Firstly, the Poisson sum formula~[4] maybe used to establish the result
$$\sum_{n = -\infty}^{\infty}\ e^{-\frac{\alpha}{2}n^2}\cos nz\ =\ {\sqrt{\frac{2\pi}{\alpha}}} \ \sum_{n = -\infty}^{\infty}\ e^{-\frac{1}{2\alpha}(2n\pi + z)^2}.$$
Using this relation, $F(z)$ may also be written as
$$F(z)\ =\ e^{\frac{iz}{2}}\ e^{-\frac{1}{2\alpha}z^2}\ \sum_{n = -\infty}^{\infty} \ e^{-\frac{2}{\alpha}n^2\pi^2}\  e^{-\frac{2}{\alpha}n\pi z}.$$
Restoring the appropriate time scales, the correlation function can then be written in either of the equivalent forms
\begin{align}
C_1(t)\ &=\ C_1(0)\ e^{\frac{it}{2\tau_b}}\ \frac{\sum_{n = -\infty}^{\infty}\ e^{-\frac{\tau_a}{2\tau_b}n^2}\ \cos\frac{nt}{\tau_b}}{\sum_{n = -\infty}^{\infty}\ e^{-\frac{\tau_a}{2\tau_b}n^2}} \ ,\\
C_1(t)\ &=\ C_1(0)\ e^{\frac{it}{2\tau_b}}\ e^{-\frac{t^2}{2\tau_a\tau_b}}\ \frac{\sum_{n = -\infty}^{\infty}\ e^{-\frac{2\tau_b}{\tau_a}n^2\pi^2}\ e^{-\frac{2t}{\tau_a}n\pi}}{\sum_{n = -\infty}^{\infty}\ e^{-\frac{2\tau_b}{\tau_a}n^2\pi^2}}\ . \label{}
\end{align}

This is a remarkable equality. In the first expression the periodic character of $C_1$ is apparent, while in the second form $C_1$ seems to exhibit gaussian decay. An explanation for this peculiar feature may be found by noting that as the sum over $n$ runs from $-\infty$ to $\infty$ it has both exponentially decaying and exponentially growing terms.

If we now consider the limit $\hbar\rightarrow 0$ we get 
$$\tau_a\ =\ \hbar\beta\rightarrow 0$$ 
and 
$$\tau_b\ =\ \frac{mR^2}{\hbar} \rightarrow\infty ,$$
but 
$${\tau_a}{\tau_b}\ =\ mR^2\beta .$$ 
Therefore, in the limit that $\hbar\rightarrow0$, the only non-vanishing term in Eq.(6) is the one with $n = 0$ and we deduce that 
$$\lim_{\hbar\rightarrow 0} C_1(t)\ =\ C_1(0)\ e^{-\frac{t^2}{2\beta mR^2}},$$
$$\lim_{\hbar\rightarrow 0} C_2(t)\ =\ C_1(0)\ e^{-\frac{t^2}{2\beta mR^2}}\ .$$

Thus in the classical limit the correlation functions $C_1$and $C_2$ are identical and exhibit gaussian decay. Atleast in this model, the strict periodicity of the quantum expression for the correlation function and the apparently irreversible decay shown by the classical expression are completely reconciled in the limit $\hbar\rightarrow 0$.

\section{CLASSICAL STATISTICAL MECHANICS}
\medskip\noindent
We now consider the same system of identical non-interacting particles moving on a circle from the point of view of classical mechanics. If the average energy at $t = 0$ is specified, then in terms of the Hamiltonian
$$H\ =\ \frac{l^2}{2mR^2},$$ 
where $l$ is the angular momentum, the density matrix is given by
$$\rho\ =\ \frac{1}{Z}\ e^{-{\beta H}},~~~~{\rm with}~~~~Z = 2\pi \ \int_{-\infty}^{\infty}\ e^{-{\beta H}}\ dl.$$
The Hamilton equations
$$\dot\phi\ =\ \frac{\partial H}{\partial l}\ =\ \frac{l}{mR^2}~~~~{\rm and}~~~~\dot l\ =\  -\ \frac{\partial H}{\partial\phi}\ =\ 0$$
lead to the solutions
$$l\ =\ {\rm constant}~~~~~{\rm and}~~~~~\phi(t)\ =\ \phi\ +\ \frac{lt}{mR^2}.$$
Hence 
$$x(t)\ =\ R\cos\phi(t)\ =\ R\cos\left(\phi\ +\ \frac{lt}{mR^2}\right).$$
The correlation function $C_1$ is then given by
$$C_1(t)\ =\ \frac{{\int_0^{2\pi}\int_{-\infty}^{\infty}\ x(0)\ x(t)\ e^{-\frac{\beta l^2}{2mR^2}}\  d\phi\ dl}}{2\pi\ \int_{-\infty}^{\infty}\ e^{-\frac{\beta l^2}{2mR^2}}\ dl}.$$
The integrals may be evaluated, with the result
$$C_1(t)\ =\ \frac{1}{2}\ R^2\ e^{-\frac{t^2}{2\beta mR^2}}.$$
This result agrees with the classical limit expressions derived earlier.

\section{CONCLUSION}
\medskip\noindent 
We have demonstrated that the position correlation functions for a system of particles moving on a circle, and having a specified mean energy, may be written in a form which exhibits their transformation properties. By using the Poisson sum formula the expressions were rewritten to enable the limit $\hbar\rightarrow 0$ to be taken. We showed finally that in that limit the correlation functions become identical to the form given by classical statistical mechanics, which exhibits gaussian decay.

\section{REFERENCES}
\bigskip
{[1]} H.~L.~Frisch, {\it Phys.~Rev.} {\bf 109} 22 (1958).

\medskip
{[2]} A.~Hobson and D.~L.~Loomis, {\it Phys.~Rev.} {\bf 173} 285 (1968).

\medskip
{[3]} E.~T.~Jaynes, {\it Papers on probability, statistics and statistical physics}, Synthese Library, Vol.158 (ed. R.~D.~Rosenkrantz), Reidel, Dordrecht (1983).

\medskip
{[4]} P.~M.~Morse and H.~Feshbach, {\it Methods of Theoretical Physics, Part I}, New York: McGraw-Hill, 466-7 (1953).

\end{document}